\documentclass[%
 aip,
 amsmath,amssymb,
 reprint,%
]{revtex4-1}

\usepackage{graphicx}
\usepackage{dcolumn}
\usepackage{bm}

\usepackage[utf8]{inputenc}
\usepackage[T1]{fontenc}
\usepackage{mathptmx}
\usepackage{etoolbox}
\usepackage{soul, color, xcolor}
\usepackage{float}
\usepackage{multirow}
\usepackage{array}
\usepackage{booktabs}
\usepackage{siunitx}
\usepackage{adjustbox}
\usepackage{makecell}
\usepackage{float}

\makeatletter
\def\@email#1#2{%
 \endgroup
 \patchcmd{\titleblock@produce}
  {\frontmatter@RRAPformat}
  {\frontmatter@RRAPformat{\produce@RRAP{*#1\href{mailto:#2}{#2}}}\frontmatter@RRAPformat}
  {}{}
}%
\makeatother
\begin{document}

\preprint{AIP/123-QED}

\title{Enhancing Accuracy and Feature Insights in Hydration Free Energy Predictions for Small Molecules with Machine Learning}
\author{Mingjun Han}
\thanks{These authors contributed equally.}
\affiliation{School of Science, Harbin Institute of Technology, Shenzhen, 518055, China}
\affiliation{Shenzhen Key Laboratory of Advanced Functional Carbon Materials Research and Comprehensive Application, Shenzhen 518055, China.} 
\author{Yukai Zhang}
\thanks{These authors contributed equally.}
\affiliation{Shenzhen Institute for Advanced Study, University of Electronic Science and Technology of China, Shenzhen 518000, China}
 
\author{Taotao Yu}
\affiliation{School of Science, Harbin Institute of Technology, Shenzhen, 518055, China}
\affiliation{Shenzhen Key Laboratory of Advanced Functional Carbon Materials Research and Comprehensive Application, Shenzhen 518055, China.} 
\author{Guodong Du}
\affiliation{School of Science, Harbin Institute of Technology, Shenzhen, 518055, China}

\author{ChiYung Yam}
\thanks{Authors to whom correspondence should be addressed: yamcy@uestc.edu.cn and denghaojian@hit.edu.cn}
\affiliation{Shenzhen Institute for Advanced Study, University of Electronic Science and Technology of China, Shenzhen 518000, China}

\author{Ho-Kin Tang}
\thanks{Authors to whom correspondence should be addressed: yamcy@uestc.edu.cn and denghaojian@hit.edu.cn}
\affiliation{School of Science, Harbin Institute of Technology, Shenzhen, 518055, China}
\affiliation{Shenzhen Key Laboratory of Advanced Functional Carbon Materials Research and Comprehensive Application, Shenzhen 518055, China.}

\date{\today}

\begin{abstract}

The accurate prediction of solvation free energy is of significant importance as it governs the behavior of solutes in solution. In this work, we apply a variety of machine learning techniques to predict and analyze the alchemical free energy of small molecules. Our methodology incorporates an ensemble of machine learning models with feature processing using the K-nearest neighbors algorithm. Two training strategies are explored: one based on experimental data, and the other based on the offset between molecular dynamics (MD) simulations and experimental measurements. The latter approach yields a substantial improvement in predictive accuracy, achieving a mean unsigned error (MUE) of 0.64 kcal/mol. Feature analysis identifies molecular geometry and topology as the most critical factors in predicting alchemical free energy, supporting the established theory that surface tension is a key determinant. Furthermore, the feature analysis of offset results highlights the relevance of charge distribution within the system, which correlates with the inaccuracies in force fields employed in MD simulations and may provide guidance for improving force field designs. These results suggest that machine learning approaches can effectively capture the complex features governing solvation free energy, offering novel pathways for enhancing predictive accuracy.
   
\end{abstract}

\maketitle
\section{INTRODUCTION}

Alchemical free energy calculations, also known as Free Energy Perturbation (FEP), play a critical role in the early stages of drug discovery by screening out potential candidates.~\cite{cournia2017relative,de2019computationally,kuhn2017prospective,georgiou2017pushing} However, their widespread application has been hindered by challenges such as high computational costs and inaccuracies in potential energy functions. Despite these limitations, there is growing interest in enhancing these methods to expand their utility. A critical aspect of FEP is the prediction of solvation free energy, which plays a vital role in understanding how solutes behave in solution. Accurately predicting solvation free energies provides valuable insights that aid in understanding molecular interactions in solution, with significant implications for drug design and materials science.~\cite{ramsundar2016democratizing} 
One important application of free energy calculations is in predicting protein-ligand binding free energies, which deepens the understanding of such binding mechanisms and provides critical theoretical support and guidance for drug design and biomedical research.~\cite{shivakumar2010prediction,martins2014prediction,mobley2009predictions} Achieving accurate predictions in this context often hinges on the optimization of force field parameterization. This remains a significant challenge, as it requires considerable time and effort to develop stable models that consistently provide accurate results. 

Recent efforts have focused on refining force field parameterization to enhance the accuracy and generality of parameter sets.\cite{tian2019ff19sb,huang2017charmm36m,slochower2019binding} However, determining the necessary modifications to improve the accuracy of parameter sets remains a challenging task, as the performance of force fields depends on multiple factors, including molecular chemical structure and coverage of conformational space. Despite significant progress in force field parameterization, classical force field-based calculations still face fundamental accuracy limitations, particularly due to the simplifications in models such as fixed-charge force fields that neglect polarization effects. The recognition of these limitations has spurred the development of post-processing methods based on quantum mechanical (QM) calculations.\cite{beierlein2011simple,konig2014predicting} These methods introduce correction terms into classical force field-based free energy predictions to correct or compensate for errors introduced by simplified models like fixed-charge force fields, aiming to improve the accuracy and reliability of free energy calculations. Therefore, current research and method development aim to leverage the strengths of both classical force field and quantum mechanical calculations to enhance the predictive accuracy of important biomolecular interactions and improve the accuracy of their physical descriptions.

In recent years, data-driven machine-learning (ML) methods have experienced a resurgence of interest in the field of drug discovery. Impressive advances have been made in areas such as quantum chemical calculations, virtual screening, and free energies calculations. In the context of predicting molecular properties, it have been used in predicting properties such as solvation free energy, protein-ligand binding affinity, and others. These studies highlight the versatility and potential of ML in improving our understanding and prediction of molecular interactions. Zhang et al. \cite{zhang2023machine} investigates the prediction of hydration free energy using ML models that do not rely on specific input features associated with molecular structure. This approach demonstrated the ability of ML to accurately predict hydration free energies without the need for atom-, bond-, or geometry-specific descriptors. Osaki et al. \cite{osaki20223d} introduced another ML approach called 3D-RISM-AI, which predicts protein-ligand binding affinity by incorporating hydration free energy as a key factor to improve the accuracy of binding free energy prediction. This study highlights the importance of hydration free energy in improving the prediction of protein-ligand interactions. In addition, Alibakhshi $\&$ Hartke~\cite{alibakhshi2021improved} presents a kernel-based ML model designed to predict the free energies of solvated organic molecules in water using implicit solvent molecular dynamics simulations. The model illustrates the potential of ML in estimating the free energy of solvation for various molecular types. Lim $\&$ Jung \cite{lim2019delfos} developed Delfos, a deep learning model for predicting the free energy of solvation in general-purpose organic solvents, which improves prediction accuracy. Meng et al. further advanced solvation free energy prediction using a $\delta$-learning approach.~\cite{meng2023something} Ansari et al. accurately predicted the solvation free energy of organic molecules from paired-atom interactions via graph-attentive networks and message-passing neural networks, which outperforms the existing methods.~\cite{ansari2021accurate} Overall, these studies demonstrate the substantial progress and potential of ML models for predicting a wide range of molecular properties, from solvation free energy to protein-ligand binding affinity. By leveraging ML techniques, researchers can improve the accuracy and efficiency of predicting molecular properties, thereby advancing the field of computational chemistry and molecular modelling.

The achievements made using ML methods are remarkable, but the methods usually have specific requirements on the training set, one of the widely used datasets is the FreeSolv database, which contains 642 molecules.~\cite{mobley2014freesolv} The limited variety of samples in this dataset can lead to data imbalance, potentially resulting in models with poor generalization ability or inaccurate predictions for certain molecular properties. Additionally, the choice of features during molecular fingerprinting significantly influences prediction accuracy. Identifying important features is crucial for optimizing the modeling process. 

In this paper, we first performed feature transformation on the molecules in the dataset, converting them into numerical features suitable for machine learning models by applying various molecular fingerprinting methods. We also utilized the K-Nearest Neighbors (KNN) algorithm to handle feature imputation. In Section II, we describe the preprocessing steps in detail, including the feature engineering and imputation process. In Section III, we explore the predictive performance of multiple ML models, including Support Vector Machine (SVM), Random Forest (RF), Multiple Linear Regression (MLR), Deep Neural Network (DNN), XGradient Boosting (XGB), and ensemble learning. We also compare our results with those from previous studies, as summarized in Table \ref{Comparison}. Among these, the first reference study utilized the MolPropsAPFP molecular fingerprint on the FreeSolv database~\cite{mobley2014freesolv}, while the second study employed the FCFP molecular fingerprint~\cite{hutchinson2019solvent}. The third reference study collected data from previous literature,\cite{engberts1998solvent} using the MolProps molecular fingerprint. We also compared the results using the interpretable Graph Interaction Network (CIGIN) as the machine learning model.\cite{pathak2021learning} This study was conducted on the Solv@TUM database\cite{hille2019generalized} and the FreeSolv dataset,\cite{mobley2014freesolv} achieving a final test error of 0.76 kcal/mol. Notably, our results yield a substantial improvement in predictive accuracy compared to these prior studies. Features Analysis focuses on the analysis of feature importance using heatmaps, providing insights into the key factors influencing solvation free energy. Our findings suggest that molecular polarizability and charge distribution are critical determinants. Finally, in Section 5, we summarize the overall work and offer prospects for future research on predicting molecular properties using ML models.
\begin{table}[H]
\centering
\caption{Comparison of the predictive effectiveness of the best-performing machine learning models from previous research efforts.\cite{scheen2020hybrid,hutchinson2019solvent,famini1999using} (with MUE as the assessment metric)}
\label{Comparison}
    \begin{adjustbox}{width=0.48\textwidth}
    \renewcommand{\arraystretch}{1.5}
    \begin{ruledtabular}
    \begin{tabular}{ccccc}
    & Ref. [25] & Ref. [21] & Ref. [26] & This Work\\ 
    \hline
    MUE & 0.76 & 1.65 & 0.91 & \textbf{0.64}\\ 
    \end{tabular}
    \end{ruledtabular}
    \end{adjustbox}
\end{table}

\section{THEORY AND METHODS}
\subsection{Data set acquisition}

The dataset used in this study is sourced from the FreeSolv database (version 0.52),~\cite{mobley2014freesolv} a widely utilized benchmark dataset for the prediction of solvation free energies. The FreeSolv database provides both experimental measurements and theoretical calculations of solvation free energies for 642 small neutral organic molecules, making it an ideal resource for building and evaluating computational models aimed at predicting solvation properties. 

Out of the 642 molecules, 47 were specifically selected for the SAMPL4 blind challenge, an influential benchmarking exercise designed to evaluate and compare different predictive models in the context of solvation and molecular interactions. These 47 molecules are designated as the test set, as their free energy values are excluded from the training process to allow for unbiased model validation. This blind testing ensures that the model’s performance can be evaluated based on unseen data, simulating a real-world predictive scenario. 

The remaining 595 molecules serve as the training set for developing and optimizing the computational models. These molecules provide a rich dataset to enable the model to learn relevant patterns and relationships between molecular structures and their solvation energies. To avoid data leakage and ensure generalizability, none of the molecules from the test set were used in the model-building phase. By carefully separating the test and training sets, this study adheres to best practices in machine learning and statistical analysis, ensuring that the reported performance metrics reflect the model's ability to generalize beyond the training data.

\subsection{feature preprocessing}
In the field of machine learning, directly utilizing three-dimensional molecular structures for prediction is challenging due to the inherent complexity of these structures, which often involves intricate spatial information and multidimensional data. Traditional ML algorithms struggle to capture these features effectively. Therefore, it is essential to convert these molecular structures into numerical features that are suitable for ML.
For these molecular samples, a variety of feature generators are available, allowing for the creation of descriptors that can be utilized both individually and in concatenated forms. The descriptors employed in this investigation include APFP, ECFP6, TOPOL, MolProps, and their concatenated combinations, MolPropsAPFP and MolPropsECFP6. Additionally, the descriptor X-NOISE is included to introduce noise into the dataset.

Furthermore, during the generation of molecular features, we observed that certain features exhibited non-numeric values. While this is reasonable, traditional ML models are unable to effectively process these features. Initially, non-numeric features are removed to ensure that only relevant numerical data is retained, which aids in effective data analysis. For molecules that lack certain features, we employ the K-Nearest Neighbor (KNN) algorithm to estimate missing values, thereby preserving the dataset's integrity. Finally, all feature data are normalized to create a consistent scale across the dataset. This comprehensive preprocessing approach results in a robust dataset, well-suited for analysis and subsequent model training.

\subsection{Training Strategy}
We propose two strategies for training models to predict the solvation free energy of molecules, denoted as $\Delta G$. The first strategy involves directly predicting experimental values based on molecular features, resulting in free energies referred to as $\Delta G_{Strategy1}$. The second strategy incorporates calculation values as additional molecular features. Here, the model predicts the offset between the computed values and the experimental values, allowing for a correction of the calculated free energies.
\begin{equation}\label{eqn-1}
  \Delta G_{\rm offset}(A) = \Delta G_{EXP}(A) - \Delta G_{FEP}(A),
\end{equation}
where $A$ represents an arbitrary sample in the database. Here, $\Delta G_{FEP}(A)$, is the calculated free energies obtained using MD simulations with the General Amber Force Field (GAFF).~\cite{van2005gromacs,wang2004development} The experimental value is denoted as $\Delta G_{EXP}(A)$. For each training set, defined by its descriptors, ML models were applied using 5-fold cross-validation. This resulted in a total population of $N=5$ trained models. Each individual model in $N$ predicts its own $\Delta G_{\rm offset}$ value. The offset estimator is defined as the arithmetic mean of these predicted offset values, with the standard deviation of the mean serving as a measure of precision.
In this approach, the ML model predicts the offset, and the final corrected hydration free energy is calculated by adding this offset to the computed values:
\begin{equation}\label{eqn-2}
  \Delta G_{Strategy2}(A) = \Delta G_{FEP}(A) + \left\langle\Delta \hat{G}_{\rm offset}(A)\right\rangle_{N},
\end{equation}
The precision of the $\Delta G_{Strategy2}(A)$ is determined by propagating the statistical errors from both the alchemical and ML terms.

Based on the strategy 1, we can directly predict the solvation free energy of molecules. The results obtained through this method maintain a high level of interpretability, as there is a relatively direct physicochemical relationship between the features and the outcomes. Additionally, the trained model exhibits strong generalization ability. In the strategy 2, instead of directly predicting the experimental value of the free energy, we correct the simulation value by predicting the difference between the experimental and simulated values. The motivation behind this approach is multifaceted: it preserves the correct trend set by the simulation calculations and avoids 'wasting' effort in regions where the calculated values are already highly accurate. However, its limitation lies in its weaker generalization ability, requiring the data sources (calculation methods) of the training and test sets to be highly consistent. In this study, we will analyze and discuss the results generated by both training strategies separately.

\subsection{Models and Optimization}
We implemented five distinct machine learning models: Support Vector Machines (SVMs), Random Forests (RFs), Deep Neural Networks (DNNs), Multiple Linear Regression (MLR), and XGBoost (XGB). These models were selected for their diverse strengths in handling different aspects of the prediction task. The specific hyperparameter search space for each model is comprehensively detailed in TABLE \ref{Hyperparameter space}, where all parameters influencing the model's performance are outlined.

To efficiently identify the optimal hyperparameter configurations for each model, we adopted a Bayesian optimization approach utilizing the SciKit-Optimize (SKOPT) library, version 0.5.2 \cite{SKOPT}. This method leverages an expected improvement acquisition function, which allows for more strategic and effective exploration of the hyperparameter space compared to traditional methods like random or grid searches. By focusing the search on promising regions of the space, Bayesian optimization helps to identify high-performing configurations with fewer iterations. We set the optimization steps to a maximum of 40, as prior experience indicates that model convergence typically occurs well before this limit.

For each iteration during the optimization process, the MUE on the validation set, averaged across cross-validation folds, was computed and returned to the SKOPT routine. This metric served as the cost function to guide the selection of the next set of hyperparameters. With each successive call, SKOPT further refined its search, aiming to minimize the cost function and achieve the best possible performance. This iterative process continued until either convergence was observed or the step limit was reached. A complete visualization of this workflow, outlining the model training and hyperparameter optimization process, is provided in Figure 1.

\begin{figure}[hbtp]
  \centering
  \includegraphics[width=0.4\textwidth]{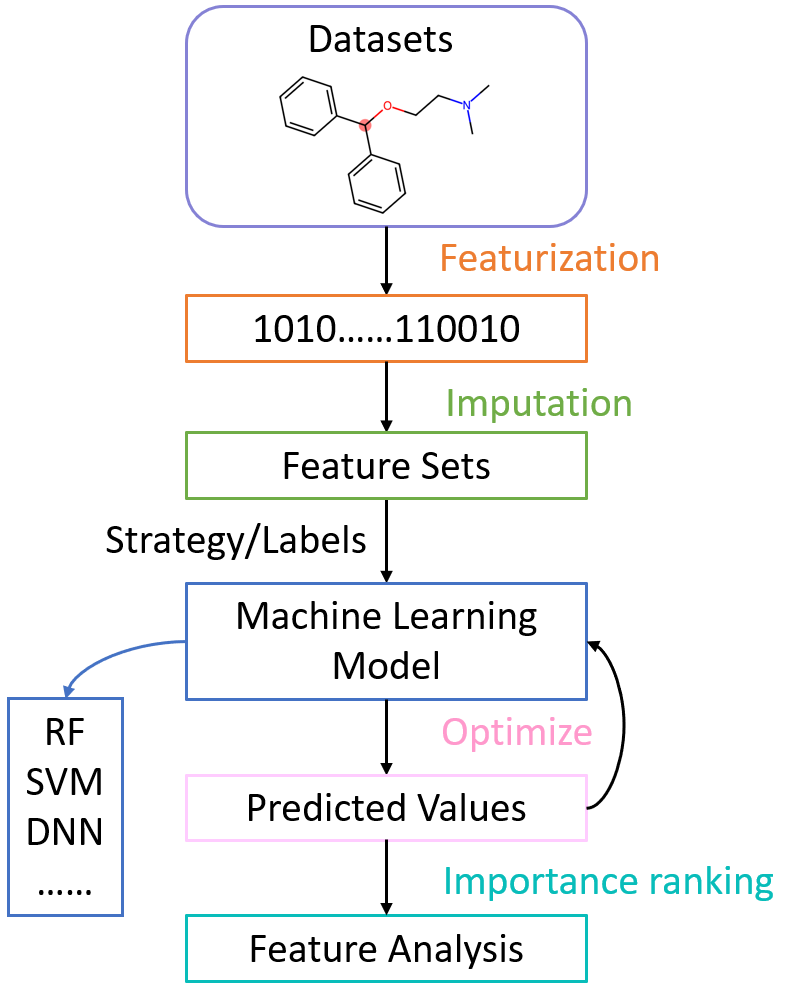}
  \caption{The workflow for this study consists of dataset acquisition, feature encoding, data cleaning, model training, optimization, prediction, and feature analysis.}
\end{figure}

\subsection{Ensemble}
For each individual model, the process begins with the identification of the optimal set of hyperparameters that minimizes the MUE during cross-validation. This step ensures that each model is fine-tuned to achieve its best possible performance on the validation data, balancing the trade-off between model complexity and accuracy. Once the optimal hyperparameters have been identified, the model is retrained using the entire training dataset. This retraining phase allows the model to fully utilize all available data, enhancing its capacity to generalize to unseen instances, which is critical for achieving robust predictive performance.

After retraining, the model is applied to the test set, where it generates predictions based on the newly trained configuration. These predictions represent the final outputs of the individual models. However, instead of relying solely on the performance of a single model, we employ an ensemble approach to further boost accuracy and reliability. 

In the ensemble strategy, predictions from all models are combined by averaging their outputs. This method leverages the unique strengths of each model, as different models may excel at capturing distinct patterns or relationships within the data. By integrating their predictions, the ensemble mitigates the potential weaknesses or biases inherent in any single model. The result is typically an overall performance improvement, as the ensemble benefits from the diversity of model predictions, leading to better generalization on new data.

Finally, the ensemble predictions are compared to the actual values from the test set ($y_{\text{test}}$) to compute the final MUE. This metric provides a comprehensive measure of the ensemble's accuracy, indicating how closely the combined predictions align with the true outcomes. The ensemble learning framework not only enhances the predictive power of the system but also helps reduce the risk of overfitting and ensures more robust performance compared to any individual model. Through this approach, we aim to achieve a higher level of reliability and accuracy in the final results.

\section{RESULTS AND DISCUSSION}
\subsection{Each individual model $\&$ Ensemble}
\begin{figure}
    \centering
    \includegraphics[width=1\linewidth]{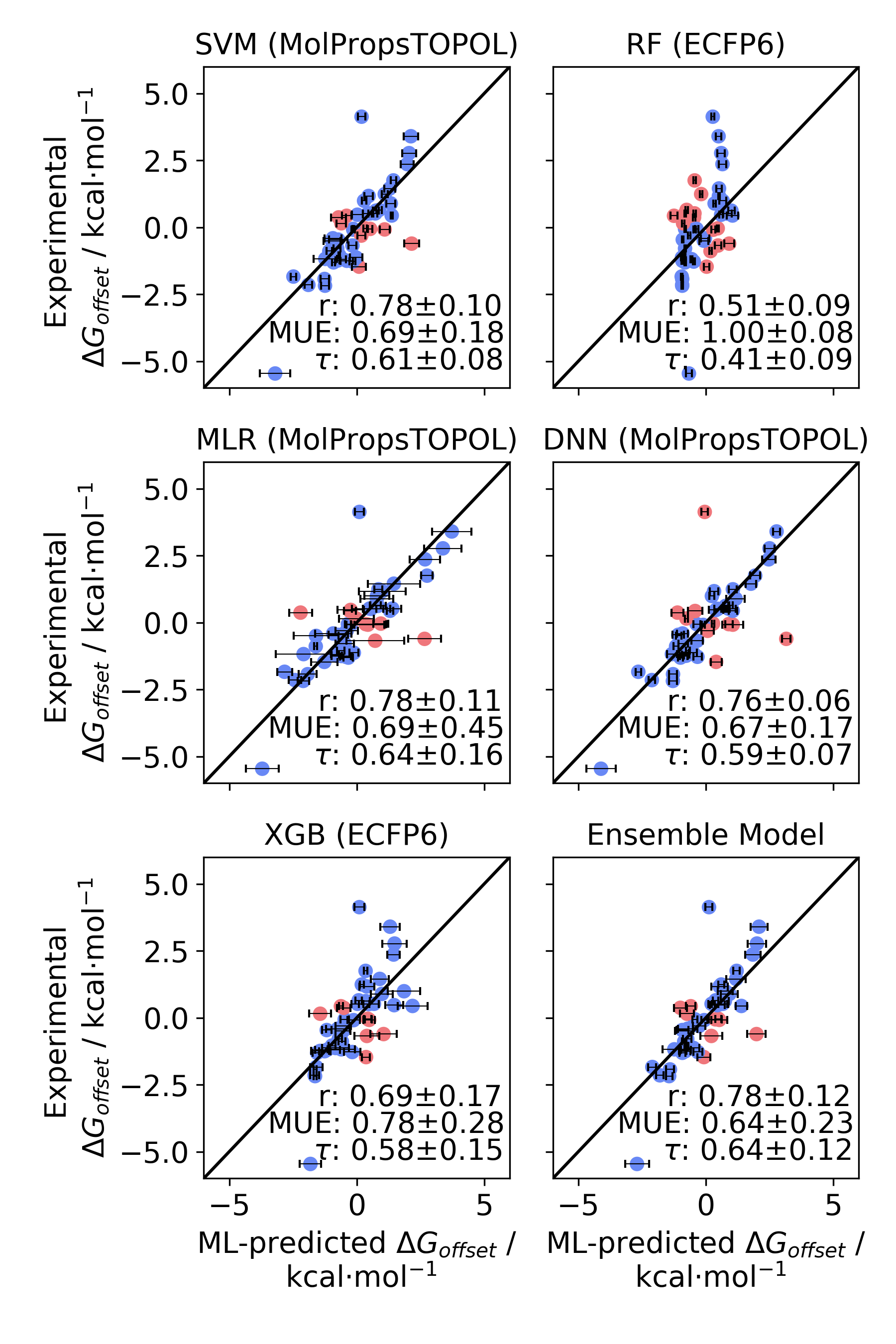}
    \caption{Comparison of the best predicted and actual values of $\Delta{G}_{\rm offset}$ by different models, where blue represents that the predicted value has the same sign as the actual value (positively corrected), red represents that the predicted value has the opposite sign to the actual value (negatively corrected), and the values on the diagonal line indicate that the experimental value is equal to the predicted value.}
    \label{fig2}
\end{figure}

Figure~\ref{fig2} presents a comparison between the predicted and actual values of $\Delta G_{\rm offset}$ for the six different models. For each model, the best performing feature set is shown. We begin by examining the results of the SVM model, a well-established supervised learning technique used for both classification and regression tasks. In regression tasks, such as Support Vector Regression (SVR), the primary objective is to minimize the error between the predicted and actual values by identifying an optimal hyperplane.
From the analysis of Figure~\ref{fig2}, the MolPropsTOPOL feature set outperforms other feature sets. This feature set not only excels in the SVM model but also demonstrates strong performance in other models, including multiple linear regression (MLR) and deep neural networks (DNN). The effectiveness of MolPropsTOPOL could be attributed to its comprehensive feature set, which likely captures essential information and enhances the ability of the model to identify critical patterns within the data. As a result, the SVM model achieves commendable performance, with a controlled MUE of approximately 0.69 and a lower relative error when compared to the actual values.

In contrast, MLR is a traditional statistical learning technique that models the linear relationship between a target variable and one or more features. In MLR, the target variable is expressed as a linear combination of the features, and the model parameters are estimated by minimizing the sum of squared errors. Compared to other ML models, MLR is highly efficient since it does not involve iterative parameter updates. Despite its straightforward approach, MLR achieves a commendable predictive performance, with a MUE of approximately 0.69, comparable to that of the SVM and DNN models. However, the higher uncertainty observed with MLR indicate underfitting, as the model struggles to capture more complex patterns within the data.

To capture these nonlinear relationships, DNNs offer a more sophisticated approach. DNNs utilize multiple hidden layers to automatically learn and extract high-level features from data. In this study, the sigmoid activation function is employed to facilitate nonlinear transformations, allowing the model to better fit complex patterns. Through backpropagation, DNNs iteratively adjust their weights to minimize prediction error. Among the models evaluated, the DNN demonstrated the best performance, achieving a lower MUE and reduced uncertainty compared to MLR. The optimal architecture for the DNN consisted of three layers, each with 50 neurons. However, similar to MLR, the DNN also faced challenges in correcting biases introduced by outliers.

The RF model, on the other hand, combines multiple decision trees trained on randomly selected data samples. This technique enhances robustness and can mitigate some issues related to individual model biases. While RF is generally less sensitive to missing features due to its random selection process, it can be difficult to interpret and may overfit when noise is present in the data. As illustrated in Figure~\ref{fig2}, RF exhibited the weakest predictive performance among all models, using ECFP6 as its input feature set. Nevertheless, RF displayed the smallest uncertainty estimates, likely due to its effective handling of feature variability.

In addition to RF, we also implemented XGBoost, another decision tree-based algorithm that excels in handling imbalanced data and adjusting weights flexibly. Utilizing a gradient boosting framework, XGBoost iteratively trains weak learners on the residuals of previous rounds to progressively minimize prediction error. As shown in Figure~\ref{fig2}, XGBoost achieves a lower MUE than RF and demonstrates better correction for outliers. However, while XGBoost performs well, its overall performance does not surpass that of the other models. Notably, the best input feature set for XGBoost was also ECFP6, consistent with the findings for RF.

After determining the best predictions for each individual model, we further enhanced our results by applying an ensemble approach. By averaging the predictions across all models, we generated a unified ensemble prediction, which was then compared to the test set to evaluate the final MUE. The ensemble predictions generally aligned more closely with the experimental results, effectively reducing the number of outliers. However, the overall prediction error exhibited a slight increase, possibly due to the averaging process diminished the strengths of each individual model.

In addition to the general trends observed, we identified two persistent outliers in the prediction results of several ML models, as shown in Figure~\ref{fig2}. Further investigation revealed that these outliers correspond to hexane-1,2,3,4,5,6-hexol and 2-hydroxybenzaldehyde. The presence of these extreme outliers significantly contributes to the larger prediction errors across the models.
Our analysis revealed that the experimental solvation free energy of hexane-1,2,3,4,5,6-hexol was determined through temperature extrapolation.\cite{riniker2017molecular,reinisch2014prediction} Meanwhile, the substantial discrepancy observed for the hydroxybenzaldehyde between its calculated and experimental free energy can be attributed to the uncertainties in assigning the partial charges during simulation.\cite{duarte2017approaches} Incorporating these data points into Stragety 2 predictions would inevitably lead to increased errors. 

\begin{figure}
    \centering
   \includegraphics[width=1.\linewidth]{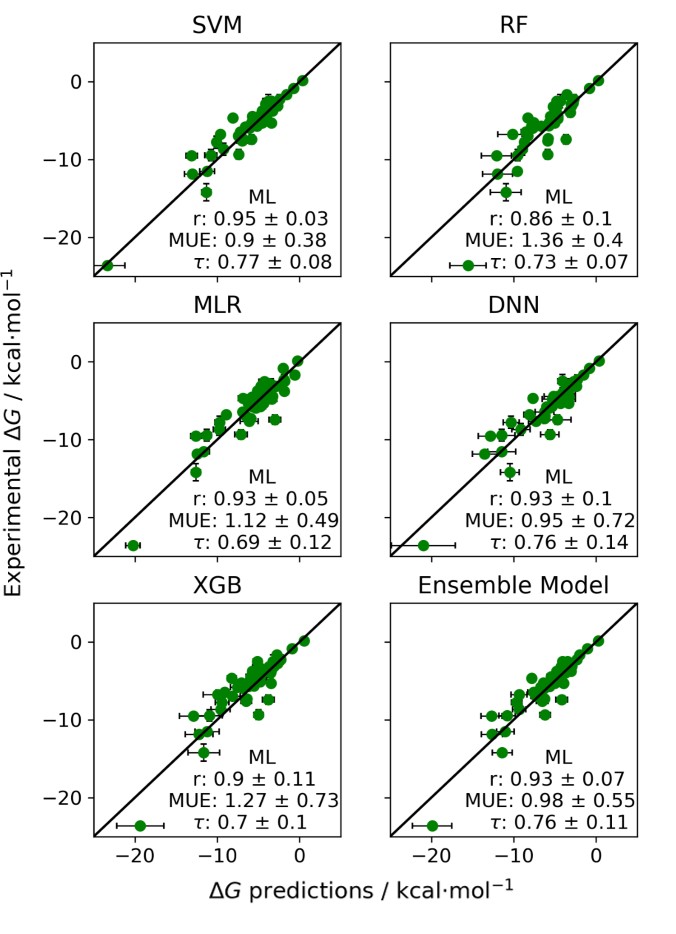}
    \caption{Prediction of Strategy 1. Each model is consistent with the models and corresponding dataset categories in Figure 2, and the Strategy 2 gives more accurate predictions compared to the Strategy 1.}
    \label{fig4}
\end{figure}

For the Strategy 1, the predictions of the models are shown in Figure~\ref{fig4}, and the corresponding feature set for each model is consistent with that in Figure~\ref{fig2}. The figure presents a comparison of predicted free energy values from five individual ML models as well as an ensemble model, against experimental values. The performance of each model is also evaluated using three key metrics: the Pearson correlation coefficient \( r \), MUE, and Kendall's tau \( \tau \). These metrics collectively assess the models' ability to predict free energy values accurately and capture the correlation between predicted and experimental outcomes. Among the models, SVM and DNN demonstrate the highest correlation with the experimental data, with \( r \)-values of \( 0.95 \pm 0.03 \) and \( 0.93 \pm 0.1 \), respectively. These high correlation values indicate that both models have a strong linear relationship with the experimental results, suggesting they are well-suited to capturing the underlying trends in the data.

In terms of prediction error, as measured by the MUE, the SVM model again performs best, achieving the lowest MUE of \( 0.9 \pm 0.38 \), closely followed by the ensemble model with an MUE of \( 0.98 \pm 0.55 \). This suggests that the SVM model makes the most accurate predictions among the individual models, while the ensemble model, which combines the predictions of all models, also performs admirably by averaging out the errors. On the other hand, the RF model exhibits the highest MUE of \( 1.36 \pm 0.4 \), indicating larger deviations from the experimental values, and its correlation coefficient is comparatively lower at \( r = 0.86 \pm 0.1 \), reflecting weaker predictive performance overall.

The ensemble model shows a well-balanced performance, with a correlation coefficient of \( r = 0.93 \pm 0.07 \), an MUE of \( 0.98 \pm 0.55 \), and Kendall's tau of \( 0.76 \pm 0.11 \). By averaging the predictions from all individual models, the ensemble approach effectively captures diverse patterns in the data, leading to more robust and accurate predictions. This method leverages the strengths of different models, mitigating the potential biases and errors inherent in any single model's predictions.

We observe that while the overall predictive performance of the model is satisfactory, there are noticeable gaps when compared to the Strategy 2. Especially for models like RF and XGB, where MUE values exceed 1 kcal/mol. Additionally, even the more accurate models, such as SVM and DNN, show greater variability in performance as indicated by relatively large error bars in the correlation coefficients and MUE. The ensemble model, though more balanced, still suffers from errors when directly predicting free energy values without leveraging simulation results as a reference.

\subsection{Comparison and Enhancement}
\begin{figure}
    \centering
    \includegraphics[width=1\linewidth]{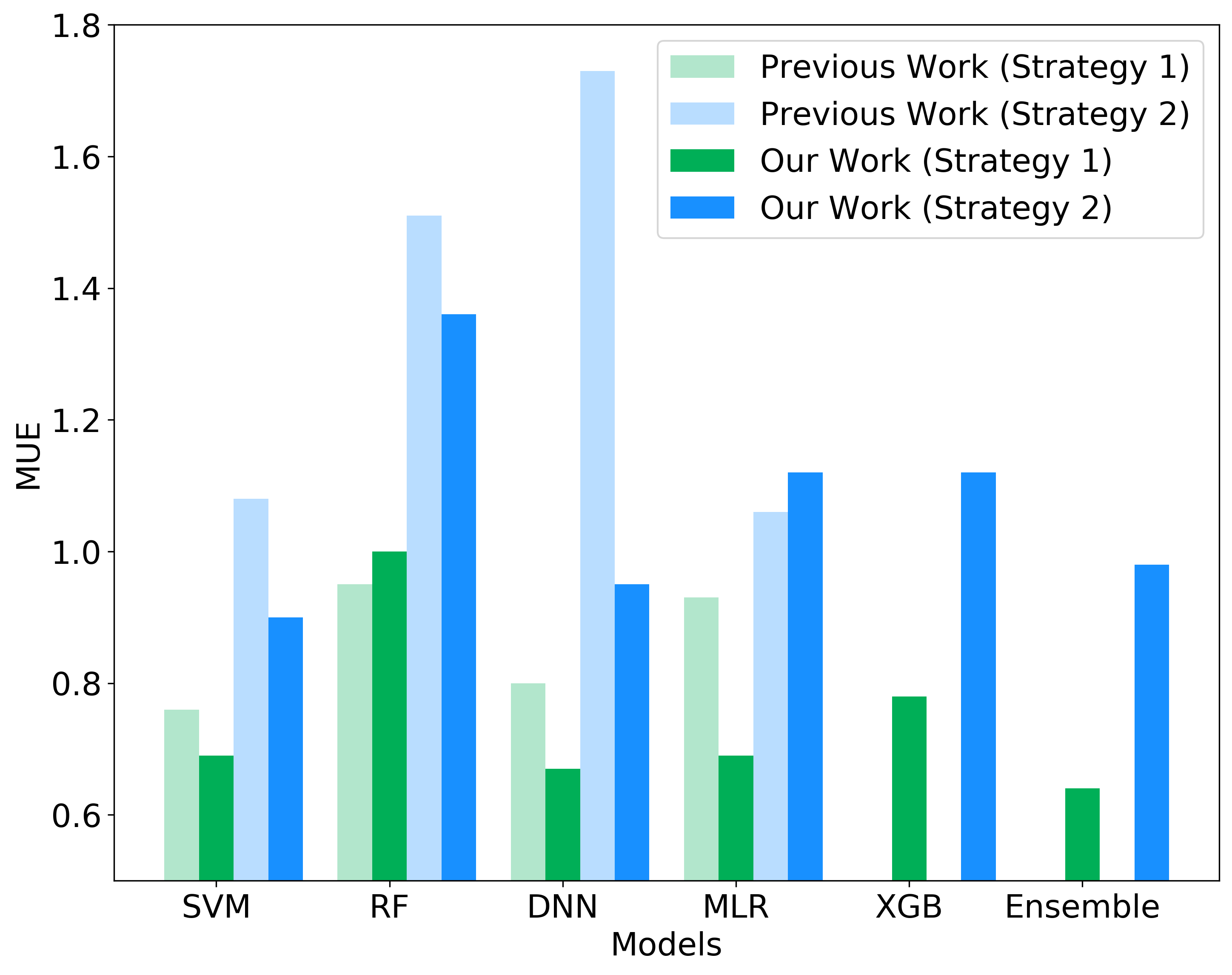}
    \caption{Comparison of model performance between Previous Work~\cite{scheen} and Our Work. Light green bars represent the MUE of models trained with Strategy 1 in Previous Work, while dark green bars represent the MUE of models trained with Strategy 1 in Our Work. Light blue and dark blue bars represent the MUE for models trained with Strategy 2 in Previous Work and Our Work, respectively. Under Strategy 1, all models show significant improvement in predictive accuracy, except for the RF model, which performs slightly worse. Under Strategy 2, all models outperform Previous Work, except for the MLR model. The Ensemble model under Strategy 1 achieves the best performance.}
    \label{fig5}
\end{figure}

We began with a comparsion with the models of previous work developed by Scheen et al.~\cite{scheen}, which utilizes four ML algorithms: SVM, RF, MLR, and DNN. Molecular features are characterized by molecular fingerprints and descriptors. To enhance model performance, we extended the original framework by adding XGBoost (XGB) and an ensemble model, as well as a new data-filling method for feature engineering. We then tested these models on the same dataset and compared their results with the models of previous work. The results indicate that, except for RF, all other models outperform the models of previous work in terms of MUE, demonstrating the effectiveness of our improvement strategy. Figure~\ref{fig5} compares the MUE for both the previous work and our work across six algorithms: SVM, RF, DNN, MLR, XGB, and an ensemble method. Overall, our models exhibit reduced MUE values compared to previous work counterparts, highlighting the success of the optimization techniques employed.

Notably, the Strategy 2 consistently outperforms the Strategy 1 across all algorithms, showing a significant difference in MUE, particularly for DNN and RF. For instance, the MUE for the DNN model of Strategy 1 is around 1.8, while the DNN model of Strategy 2 achieves approximately 1.2. This trend is also evident in other models, including RF and XGB. Following improvements, the models based on Strategy 2 retain a distinct advantage, achieving lower MUEs across all algorithms, with the enhanced Ensemble model delivering the lowest MUE overall. This indicates that while ML techniques benefit from enhancements in feature and model architecture, Strategy 2 remains more robust for this application. Future research into hybrid approaches or advanced feature extraction methods may help narrow the gap between the two strategies.

\subsection{Features Analysis}
\subsubsection{Relationships between models and datasets}

\begin{table*}[htbp]
\centering
    \caption{\label{tab:table2}Comparison of the performance of five machine learning models (RF, DNN, MLR, SVM, and XGB) using different feature sets.}
    \begin{ruledtabular}
    \begin{tabular}{cccccccccc}
    \multicolumn{2}{c}{RF} & \multicolumn{2}{c}{DNN} & \multicolumn{2}{c}{MLR} & \multicolumn{2}{c}{SVM} & \multicolumn{2}{c}{XGB} \\ 
    \midrule
    \textbf{Feature set} & \textbf{MUE} & \textbf{Feature set} & \textbf{MUE} & \textbf{Feature set} & \textbf{MUE} & \textbf{Feature set} & \textbf{MUE} & \textbf{Feature set} & \textbf{MUE}\\ 
    \midrule
    ECFP6              & 1.0$\pm$0.08 & \makecell[c]{MolProps\\TOPOL}  & 0.67$\pm$0.17  & \makecell[c]{MolProps\\TOPOL} & 0.69$\pm$0.45   & \makecell[c]{MolProps\\TOPOL} & 0.69$\pm$0.18  & ECFP6 & 0.78$\pm$0.28  \\
    \makecell[c]{MolProps\\TOPOL}     & 1.03$\pm$0.09 & TOPOL & 0.69$\pm$0.25  & \makecell[c]{MolProps\\ECFP6} & 0.88$\pm$0.29  & TOPOL & 0.72$\pm$0.19  & TOPOL & 0.81$\pm$0.23  \\
    \makecell[c]{MolProps\\APFP}      & 1.05$\pm$0.11 & \makecell[c]{MolProps\\ECFP6} & 0.74$\pm$0.19 & \makecell[c]{MolProps\\APFP}  & 1.7$\pm$0.63  & MolProps & 0.76$\pm$0.19  & \makecell[c]{MolProps\\TOPOL} & 0.83$\pm$0.25  \\
    MolProps           & 1.06$\pm$0.11 & \makecell[c]{MolProps\\APFP}  & 0.81$\pm$0.24  & MolProps & 1.72$\pm$0.75  & \makecell[c]{MolProps\\ECFP6} & 0.81$\pm$0.12  & \makecell[c]{MolProps\\ECFP6} & 0.87$\pm$0.27  \\
    \makecell[c]{MolProps\\ECFP6}     & 1.06$\pm$0.12 & ECFP6 & 0.88$\pm$0.2   & X-NOISE & 1.74$\pm$0.34  & \makecell[c]{MolProps\\APFP}  & 0.81$\pm$0.16  & MolProps & 0.9$\pm$0.26   \\
    TOPOL              & 1.11$\pm$0.2  & MolProps & 1.04$\pm$0.34  & APFP & \textbackslash & ECFP6& 0.9$\pm$0.15 & \makecell[c]{MolProps\\APFP}   & 0.94$\pm$0.23  \\
    APFP               & 1.13$\pm$0.17 & APFP & 1.11$\pm$0.25  & TOPOL & \textbackslash & APFP & 1.04$\pm$0.21  & APFP & 1.05$\pm$0.28  \\
    X-NOISE            & 1.16$\pm$0.13 & X-NOISE & 1.17$\pm$0.05  & ECFP6 & \textbackslash & X-NOISE & 1.18$\pm$0.03  & X-NOISE & 1.13$\pm$0.26  \\
    \end{tabular}
    \end{ruledtabular}
\end{table*}

The evaluation of different ML models across various feature sets reveals a significant interplay between model performance and the nature of the features used, as we compared the MUE in Table II,  the performance metrics of all molecular fingerprints under all models can be found in Appendix A, Table A1, A2. Notably, RF demonstrated consistent performance across all feature sets, achieving the lowest MUE of $1.0 \pm 0.08$ with the ECFP6 feature set. This suggests that RF is particularly well-suited for handling discrete variables, such as those in molecular fingerprints. However, its performance slightly declined when applied to the MolProps feature set, which consists of continuous variables, likely due to challenges in managing high-dimensional continuous data. In contrast, the DNN model excelled with a combination of MolProps and TOPOL feature sets, achieving the best performance with an MUE of $0.67 \pm 0.17$. This success can be attributed to the ability of DNN to navigate complex, high-dimensional feature space, making it particularly effective with continuous data. However, while its performance with discrete features remained robust, it was relatively less effective, underscoring its advantage with numerical data.

On the other hand, the MLR model exhibited less stability, particularly with discrete feature sets, showing a significant increase in error when using the MolPropsAPFP set (MUE = $1.7 \pm 0.63$). The linear nature of MLR limits its ability to capture nonlinear relationships, making it less effective with the complex interactions present in continuous data, although it performed reasonably well with the MolProps set.
Similarly, the SVM model showed strong performance with discrete feature sets like ECFP6 and TOPOL (MUE = $0.72 \pm 0.19$ and $0.69 \pm 0.18$, respectively). However, it shows variability when handling continuous variables, potentially due to its reliance on finding optimal decision boundaries in high-dimensional spaces. This variability highlights the importance of feature selection in aligning with the strengths of model.
Lastly, XGB performed well with discrete features (MUE = $0.78 \pm 0.28$ for ECFP6) but its performance deteriorated with the noisy X-NOISE feature set (MUE = $1.13 \pm 0.26$), indicating the sensitivity to data noise despite its robustness with large-scale discrete and continuous variables. This sensitivity further emphasizes the need for careful feature selection across all models.

Overall, these findings underscore the importance of aligning feature types with the strengths of the selected machine learning models. Specifically, DNN and XGB excel with continuous variables, while RF and SVM demonstrate greater consistency with discrete variables. This highlights the necessity of thoughtful feature selection that considers both the physical meaning of the features and the inherent capabilities of the models to achieve optimal predictive performance.

\subsubsection{Relationship between features and predicted outcomes}
\begin{figure}
    \centering
    \includegraphics[width=1\linewidth]{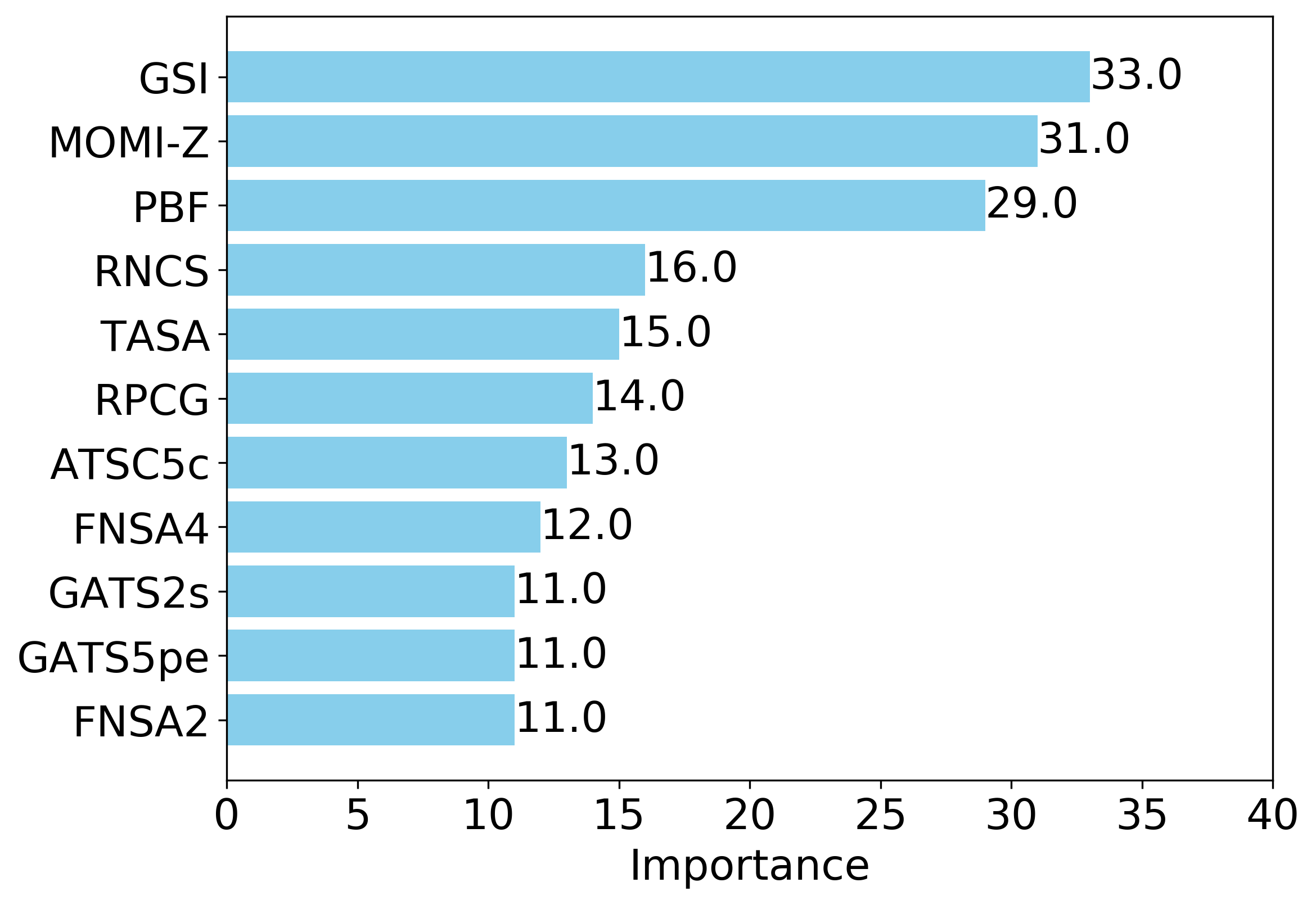}
    \caption{Importance ranking of features based on the MolPropsTOPOL dataset given by the XGB(Strategy 1) model (top 11). A higher score indicates that the feature has a greater impact on the target predictive value during training, i.e., the feature has a greater hybrid free energy correlation with the sample. The bars are labeled with the importance scores of the corresponding features. GSI stands for GeomShapeIndex.}
    \label{fig6}
\end{figure}
Next, we aim to illustrate how the hydration free energy of molecules is related to specific chemical properties, particularly from the perspective of feature importance.
For assessing feature importance, XGBoost provides an intuitive and interpretable framework. At each split node, importance scores for features are calculated, allowing us to identify which features significantly influence the model's predictions. As illustrated in Figure~\ref{fig6}, the top 11 features for the XGB(Strategy 1) model, which utilizes hydration free energies as labels, include GeomShapeIndex, MOMI-Z, PBF, RNCS, TASA, RPCG, ATSC5c, FNSA4, and others. These features are intricately linked to molecular geometry, topology, and charge distribution, all of which play crucial roles in determining molecular interactions in solution.
Here, we select the top three features with the highest scores and explain their relevance to solvation free energy. First, the geometric shape index of a molecule captures its shape characteristics, influencing interactions with solvent molecules and affecting the solvation free energy. The second feature, MOMI-Z, refers to the moment of inertia along the Z-axis. This feature reflects both the molecular geometry and the distribution of polarity, impacting the way the molecule interacts with the solvent. The above features are both related to the geometrical properties of the molecule. In general, the solvation free energy is considered as the balance between the solute-solvent interactions and the energy costs associated with increased surface tension of the solvent.~\cite{jakoby2022surface,10.1093/oso/9780198558842.003.0002} Therefore, the geometrical properties of molecule, which directly affect its solvent-accessible surface area, contribute significantly to the solvation free energy. Lastly, under Strategy 1, the PBF feature is associated with the degree of polarity. The higher the molecular polarity, the more likely it is to engage in induced dipole-dipole interactions with water molecules, and it can also facilitate more effective hydrogen bond formation. Therefore, fundamentally, we find that molecular structural characteristics and polarity have a significant impact on solvation free energy. Similarly, features related to surface areas and charge distributions, such as RNCS and TASA, provide insight into how these molecules interact with solvents. We found that the above features are primarily related to the structural characteristics and polarized charges of molecules. Given the complexity and significance of these features, detailed explanations regarding their meanings and implications are provided in the Appendix C.  This additional information will enhance understanding of how each feature contributes to the predictive performance of the models regarding hydration free energy.

\begin{figure}
    \centering
    \includegraphics[width=1\linewidth]{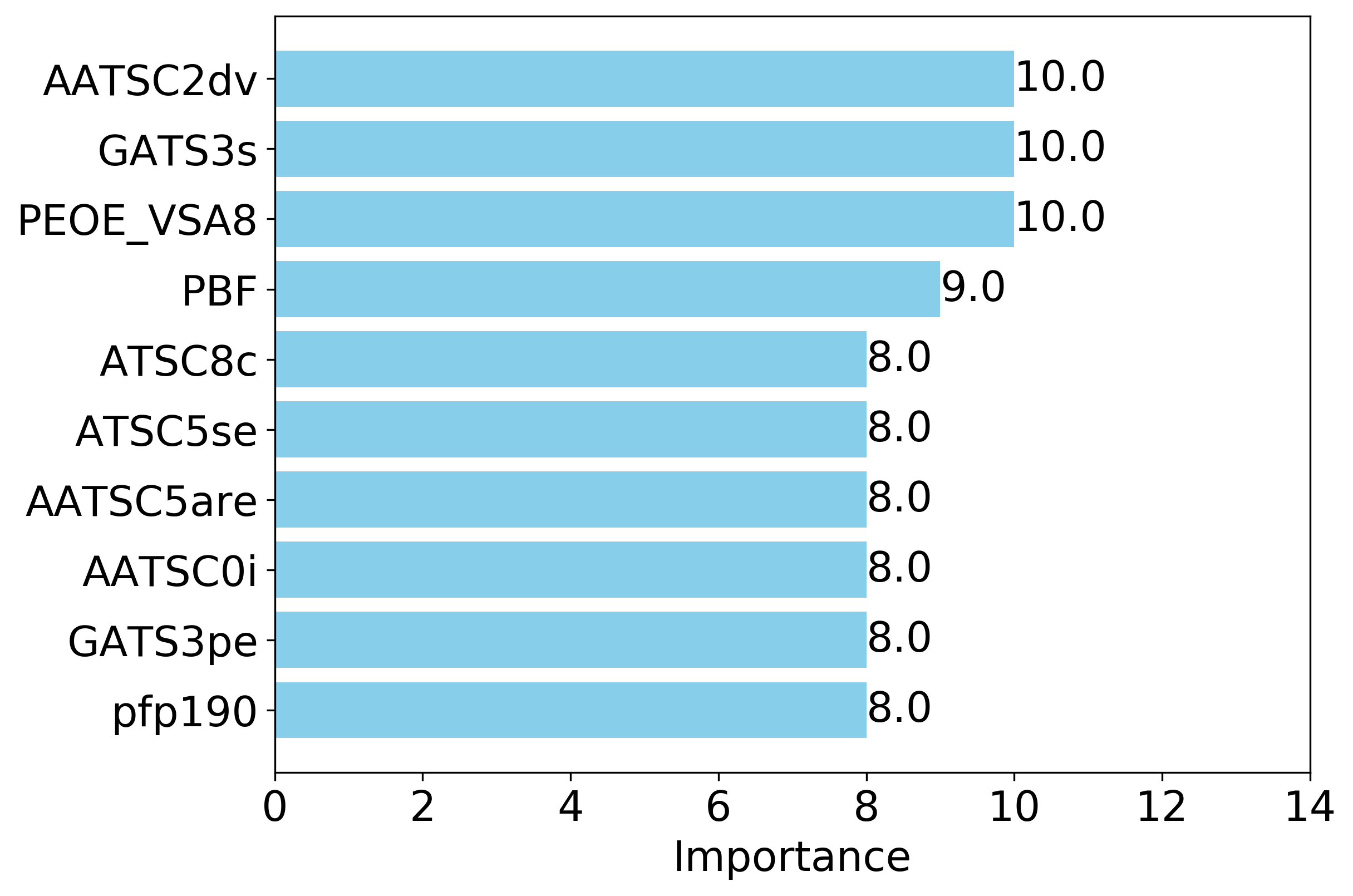}
    \caption{Importance ranking of features based on the MolPropsTOPOL dataset given by the XGB(Strategy 2) model (TOP 10).}
    \label{fig7}
\end{figure}
Building on the discussion of feature importance, the model based on Strategy 2 utilizes a dataset where the labels represent the offsets between computed and experimental values. Consequently, high-scoring features may not directly correlate with physicochemical properties closely tied to hydration free energy. Instead, these features significantly influence the offset by encompassing various molecular properties, such as topology, charge, polarity, and aromaticity, as shown in Figure~\ref{fig7}, which shows the top 10 feature importance rankings given by the XGB(Strategy 2) model based on the MolPropsTOPOL dataset. The top three features in Figure 6 are AATSC2dv, GATS3s, and PEOE-VSA8. The AATSC2dv feature is related to the structural characteristics of the molecule, while the latter two features are associated with the molecular polarity. This suggests that machine learning has captured the limitations of force field in describing polarization effects of molecules. These issues have also been the focus of ongoing efforts in MD over the past few years.\cite{tian2019ff19sb, huang2017charmm36m, slochower2019binding, jing2019polarizable} 
This aligns with our analysis from Strategy 1, indicating that for solvation free energy, both structural characteristics and molecular polarity play a dominant role. Specific definition of these features are also provided in the Appendix C.
This relationship can be understood in the context of the common discrepancies observed between simulation results and actual experimental outcomes. These deviations often arise due to the inherent limitations of current methods. In this scenario, ML techniques can serve to bridge this gap, enhancing simulation accuracy and providing valuable support for improving the precision of computational predictions.

\begin{figure}
    \centering
    \includegraphics[width=1\linewidth]{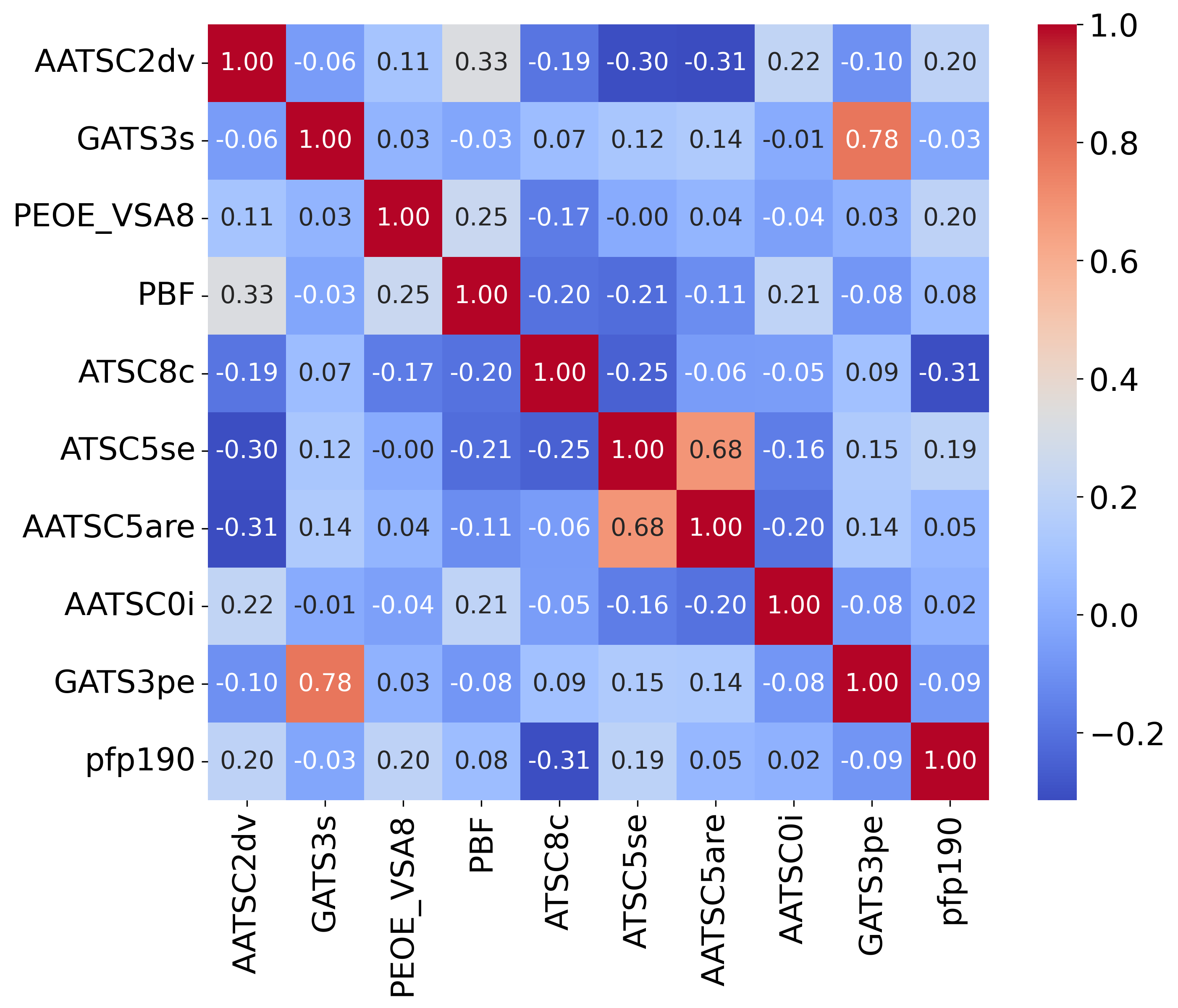}
    \caption{Correlation Heatmap of TOP 10 Features for XGB(Strategy 2) (MolPropsTOPOL). Blue color represents a negative correlation between pairs of features and red color represents a positive correlation between features. The darker the color, the higher the correlation, and the corresponding value is marked on the graph.}
    \label{fig9}
\end{figure}

To provide further insights into the interrelationships among these molecular descriptors, Figure~\ref{fig9} depicts the correlation heatmap of the top 10 features selected by the XGB(Strategy 2) model based on MolPropsTOPOL dataset.
By analyzing these correlations, we can deepen our understanding of how individual features may collectively influence the predictions, as well as identify potential redundancies or complementary aspects. As shown in Figure~\ref{fig9}, most features exhibit low correlation coefficients, generally ranging between -0.2 and 0.2. The results indicate that the selected features capture distinct molecular characteristics with minimal overlap. However, there are notable positive correlations among certain features, ATSC5se $\&$ AATSC5are (0.68), such as GATS3s $\&$ GATS3pe (0.78).

The positive correlation suggests that ATSC5se and AATSC5are are capturing similar aspects of molecular properties, particularly regarding polarity and charge distribution. Specifically, ATSC5se is an autocorrelation descriptor based on electronegativity, while AATSC5are pertains to the aromaticity and radius eigenvalue
weighting. Both features indicate regions of high or low polarizing intensity on the molecular surface that directly influence solvation properties. Similarly, GATS3s and GATS3pe also capture comparable features of the molecular electronic environment, as both relate to the molecule’s polarity and its capacity to interact with solvent molecules. Overall, these features provide valuable information captured in the training of ML models, complementing the limitations of traditional MD simulations.

\begin{figure}[htbp]
    \centering
    \includegraphics[width=1\linewidth]{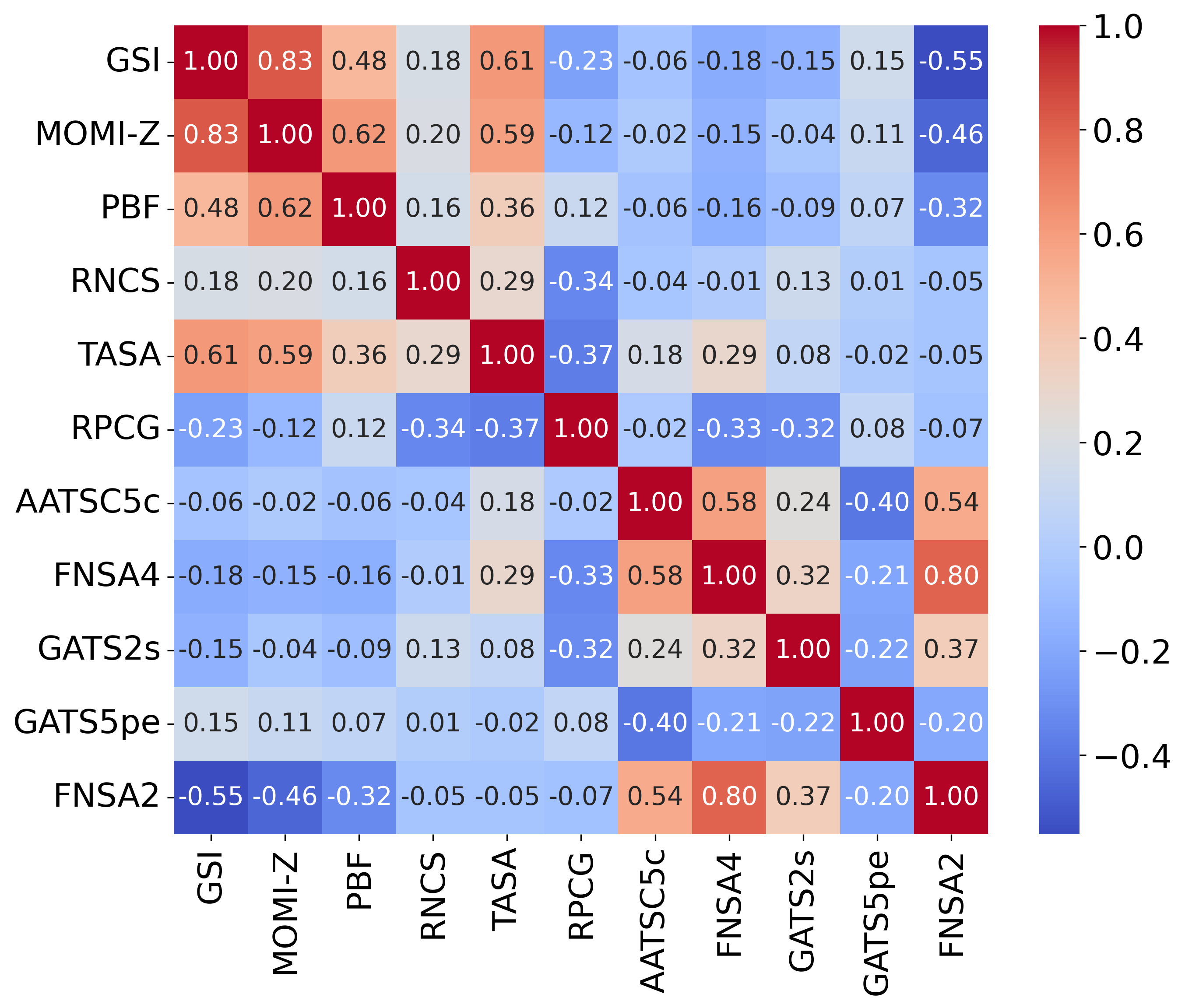}
    \caption{Correlation Heatmap of TOP 11 Features for XGB(Strategy 1). Blue color represents a negative correlation between pairs of features and red color represents a positive correlation between features. The darker the color, the higher the correlation, and the corresponding value is marked on the graph. 'GSI' stands for 'GeomShapeIndex'.}
    \label{fig8}
\end{figure}

Additionally, some features display moderate negative correlations, providing insights into inverse relationships. For instance, the negative correlation between the distance
weighted by van der Waals volume (AATSC2dv) and overall characteristics of the internal charge distribution of a molecule (ATSC8c) suggests that molecules with greater van der Waals volume tend to have weaker interactions to solvents. This may imply that larger molecules in a solvent occupy more space and therefore need to displace more solvent molecules. This creates stronger repulsive forces, resulting in a greater energy requirement to maintain the presence of these molecules, thereby affecting their solvation behavior. Similarly, the relationship between AATSC2dv and ATSC5se,  AATSC2dv and AATSC5 also suggest that greater van der Waals volume may influence overall solvation dynamics. Similar analysis is done for Strategy 1, as shown in Figure~\ref{fig8}. Overall, Figure~\ref{fig8} shows a stronger correlations among the top features, highlighting the crucial role of molecular geometries in shaping solute-solvent interactions. Additionally, this indicates the presence of non-linear relationships among these features, which collectively influence the solvation free energy.

\section{CONCLUSION}

In this work, we employed ML methods to improve the accuracy of prediction of alchemical free energy of small molecules. To achieve this goal, we trained five popular models (SVM, MLR, RF, DNN, XGB) using two distinct training strategies and comparing the test results. The results demonstrated that our ML approach outperformed previous studies, with models trained using Strategy 2 exhibiting superior overall performance compared to those trained with Strategy 1. Notably, the best-performing individual model achieved an evaluation metric (MUE) of 0.67. Building on this, we combined the individual models into an ensemble, which further improved the prediction accuracy to 0.64.

The accuracy enhancement is mainly attributed by the improvement we made in data preprocessing process and the use of model ensembling technique. For samples with missing molecular features, we utilized the KNN algorithm for feature imputation, effectively minimizing information loss that might have resulted from discarding incomplete data. Moreover, we filtered out molecules for which computational methods in the database were deemed unreliable, ensuring the robustness and reliability of our training data.

Our data analysis involved a comprehensive investigation of feature importance and model interpretability. By employing the XGB algorithm, we ranked feature importance scores under both training strategies, identifying key features that influence predictions, through the analysis of these key features, we found that they primarily pertain to the molecular structural characteristics and the molecular polarity. These aspects align with the recent advancements in molecular dynamics simulations, indicating that our model has the capability to capture these correlations effectively. Furthermore, we visualized the correlations among high-scoring features, which not only aids in effective feature selection but also enhances computational efficiency.

We evaluated the heatmaps of Strategy 1 and Strategy 2 systematically. We found that the high-scoring, highly correlated features are closely related to the molecular structural characteristics and polarity charges, which can be explained by the surface tension of the solvent and intramolecular polarization effects. Our work enriches the understanding of solvation free energy, provides insights into the features and results obtained from machine learning predictions, and offers perspectives and guidance for future research.

\section*{ACKNOWLEDGMENTS}
This study was financially supported by the National Natural Science Foundation of China (Grant No. 22073007, 12204130), Shenzhen Basic Research Key Project Fund (Grant No. JCYJ20220818103200001), Shenzhen Start-Up Research Funds (Grant No. HA11409065), and Shenzhen Key Laboratory of Advanced Functional Carbon Materials Research and Comprehensive Application (Grant No. ZDSYS20220527171407017).
\section*{AUTHOR DECLARATIONS}
\subsection*{Conflict of Interest}
The authors have no conflicts of interest to disclose.
\subsection*{Author Contributions}
All authors made significant contributions to this research. Specifically, co-first authors Mingjun Han and Yukai Zhang jointly designed the experiments, collected and analyzed the data, and drafted the initial manuscript. Author Mingjun Han was responsible for the preliminary data processing and visualization, while Author Yukai Zhang focused on the literature review and theoretical analysis. Additionally, authors Taotao Yu and Guodong Du contributed to the project by assisting with data collection and providing feedback during the study. Corresponding authors Ho-Kin Tang and ChiYung Yam provided guidance on the overall direction of the study and participated in discussions of the results and revisions of the final manuscript. All authors reviewed and approved the final version for submission.
\section*{DATA AVAILABILITY}
The data that support the findings of this study are available
from the corresponding author upon reasonable request.

\appendix
\clearpage

\section{Performance metrics for different models}
\begin{table*}[htb]
\nopagebreak
    \renewcommand{\thetable}{A1}
    \centering
    \caption{Performance metrics for different models (Part I)}
    \begin{adjustbox}{width=\textwidth}
    \begin{tabular}{cccccccc}
        \toprule
        \textbf{Model} & \textbf{Featureset} & \textbf{Labels} & \textbf{Pearson r} & \textbf{MUE} & \textbf{RMSE} & \textbf{Spearman rho} & \textbf{Kendall tau} \\
        \midrule
        XGB  & ECFP6          & G\_offset & 0.96±0.02  & 0.78±0.28  & 1.31±0.07   & 0.92±0.03    & 0.80±0.06    \\
        XGB  & TOPOL          & G\_offset & 0.96±0.02  & 0.81±0.23  & 1.21±0.08   & 0.92±0.03    & 0.79±0.04    \\
        XGB  & MolPropsTOPOL  & G\_offset & 0.95±0.02  & 0.83±0.25  & 1.45±0.01   & 0.94±0.02    & 0.82±0.04    \\
        XGB  & MolPropsECFP6  & G\_offset & 0.95±0.02  & 0.87±0.27  & 1.48±-0.00  & 0.94±0.04    & 0.81±0.04    \\
        XGB  & MolProps       & G\_offset & 0.95±0.02  & 0.90±0.26  & 1.68±-0.15  & 0.93±0.02    & 0.81±0.03    \\
        XGB  & MolPropsAPFP   & G\_offset & 0.95±0.02  & 0.94±0.23  & 1.77±-0.22  & 0.94±0.03    & 0.82±0.04    \\
        XGB  & APFP           & G\_offset & 0.93±0.03  & 1.05±0.28  & 2.37±-0.56  & 0.88±0.04    & 0.73±0.06    \\
        XGB  & X-NOISE        & G\_offset & 0.92±0.02  & 1.13±0.26  & 2.43±-0.66  & 0.88±0.05    & 0.74±0.06    \\
        XGB  & MolPropsECFP6  & G\_FEP    & 0.90±0.11  & 1.27±0.73  & 2.99±-0.46  & 0.84±0.09    & 0.70±0.10    \\
        XGB  & MolPropsAPFP   & G\_FEP    & 0.89±0.09  & 1.28±0.66  & 3.15±-0.72  & 0.86±0.11    & 0.72±0.14    \\
        XGB  & MolProps       & G\_FEP    & 0.91±0.09  & 1.30±0.70  & 2.96±-0.54  & 0.86±0.12    & 0.70±0.13    \\
        XGB  & MolPropsTOPOL  & G\_FEP    & 0.88±0.09  & 1.32±0.59  & 3.42±-0.99  & 0.85±0.10    & 0.70±0.11    \\
        XGB  & APFP           & G\_FEP    & 0.86±0.21  & 1.39±1.11  & 3.78±-0.75  & 0.72±0.36    & 0.56±0.30    \\
        XGB  & TOPOL          & G\_FEP    & 0.81±0.26  & 1.54±1.19  & 5.20±-1.61  & 0.70±0.14    & 0.55±0.14    \\
        XGB  & ECFP6          & G\_FEP    & 0.73±0.31  & 1.83±1.37  & 6.89±-2.91  & 0.68±0.32    & 0.49±0.24    \\
        XGB  & X-NOISE        & G\_FEP    & -0.07±0.25 & 3.00±1.18  & 21.44±-16.03 & -0.04±0.35   & -0.04±0.19   \\
        SVM  & MolPropsTOPOL  & G\_offset & 0.97±0.01  & 0.69±0.18  & 1.00±0.16   & 0.94±0.02    & 0.83±0.03    \\
        SVM  & TOPOL          & G\_offset & 0.96±0.01  & 0.72±0.19  & 1.09±0.12   & 0.91±0.03    & 0.79±0.05    \\
        SVM  & MolProps       & G\_offset & 0.95±0.02  & 0.76±0.19  & 1.51±-0.08  & 0.93±0.04    & 0.81±0.06    \\
        SVM  & MolPropsECFP6  & G\_offset & 0.95±0.01  & 0.81±0.12  & 1.47±-0.15  & 0.94±0.01    & 0.82±0.02    \\
        SVM  & MolPropsAPFP   & G\_offset & 0.95±0.01  & 0.81±0.16  & 1.59±-0.18  & 0.94±0.01    & 0.81±0.02    \\
        SVM  & ECFP6          & G\_offset & 0.95±0.01  & 0.90±0.15  & 1.53±-0.16  & 0.92±0.01    & 0.78±0.01    \\
        SVM  & APFP           & G\_offset & 0.94±0.01  & 1.04±0.21  & 2.20±-0.55  & 0.92±0.02    & 0.78±0.03    \\
        SVM  & X-NOISE        & G\_offset & 0.92±0.00  & 1.18±0.03  & 2.61±-0.97  & 0.86±0.00    & 0.72±0.00    \\
        SVM  & MolProps       & G\_FEP    & 0.95±0.03  & 0.90±0.38  & 1.58±0.02   & 0.91±0.06    & 0.77±0.08    \\
        SVM & TOPOL & G\_offset & 0.96±0.01 & 0.72±0.19 & 1.09±0.12 & 0.91±0.03 & 0.79±0.05 \\ 
        SVM & MolProps & G\_offset & 0.95±0.02 & 0.76±0.19 & 1.51±-0.08 & 0.93±0.04 & 0.81±0.06 \\ 
        SVM & MolPropsECFP6 & G\_offset & 0.95±0.01 & 0.81±0.12 & 1.47±-0.15 & 0.94±0.01 & 0.82±0.02 \\ 
        SVM & MolPropsAPFP & G\_offset & 0.95±0.01 & 0.81±0.16 & 1.59±-0.18 & 0.94±0.01 & 0.81±0.02 \\ 
        SVM & ECFP6 & G\_offset & 0.95±0.01 & 0.90±0.15 & 1.53±-0.16 & 0.92±0.01 & 0.78±0.01 \\ 
        SVM & APFP & G\_offset & 0.94±0.01 & 1.04±0.21 & 2.2±-0.55 & 0.92±0.02 & 0.78±0.03 \\ 
        SVM & X-NOISE & G\_offset & 0.92±0.0 & 1.18±0.03 & 2.61±-0.97 & 0.86±0.0 & 0.72±0.0 \\ 
        SVM & MolProps & G\_FEP & 0.95±0.03 & 0.90±0.38 & 1.58±0.02 & 0.91±0.06 & 0.77±0.08 \\ 
        SVM & MolPropsAPFP & G\_FEP & 0.94±0.02 & 0.93±0.32 & 1.85±-0.18 & 0.91±0.04 & 0.77±0.06 \\ 
        SVM & MolPropsTOPOL & G\_FEP & 0.93±0.04 & 1.03±0.34 & 2.27±-0.45 & 0.87±0.04 & 0.71±0.05 \\ 
        SVM & MolPropsECFP6 & G\_FEP & 0.92±0.04 & 1.12±0.24 & 2.89±-0.9 & 0.89±0.05 & 0.72±0.06 \\ 
        SVM & TOPOL & G\_FEP & 0.78±0.12 & 1.55±0.68 & 6.02±-3.02 & 0.76±0.07 & 0.60±0.08 \\ 
        SVM & APFP & G\_FEP & 0.85±0.10 & 1.63±0.87 & 4.95±-1.93 & 0.71±0.08 & 0.53±0.08 \\ 
        SVM & ECFP6 & G\_FEP & 0.69±0.13 & 1.86±0.72 & 8.13±-4.65 & 0.69±0.10 & 0.51±0.11 \\ 
        SVM & X-NOISE & G\_FEP & -0.25±0.27 & 2.93±0.14 & 20.5±-15.89 & -0.27±0.33 & -0.18±0.24 \\ 
        \bottomrule
    \end{tabular}
    \end{adjustbox}
\end{table*}
\begin{table*}
    \renewcommand{\thetable}{A2}
    \centering
    \caption{Performance metrics for different models(Part II)}
    \begin{adjustbox}{width=\textwidth}
    \begin{tabular}{cccccccc}
        \toprule
        \textbf{Model} & \textbf{Featureset} & \textbf{Labels} & \textbf{Pearson r} & \textbf{MUE} & \textbf{RMSE} & \textbf{Spearman rho} & \textbf{Kendall tau} \\
        \midrule
        RF & ECFP6 & G\_offset & 0.94±0.01 & 1.00±0.08 & 1.87±-0.44 & 0.89±0.00 & 0.75±0.00 \\ 
        RF & MolPropsTOPOL & G\_offset & 0.94±0.01 & 1.03±0.09 & 1.94±-0.44 & 0.91±0.01 & 0.78±0.02 \\ 
        RF & MolPropsAPFP & G\_offset & 0.93±0.01 & 1.05±0.11 & 2.01±-0.48 & 0.91±0.02 & 0.77±0.02 \\ 
        RF & MolProps & G\_offset & 0.93±0.01 & 1.06±0.11 & 2.07±-0.50 & 0.90±0.01 & 0.76±0.01 \\ 
        RF & MolPropsECFP6 & G\_offset & 0.93±0.01 & 1.06±0.12 & 2.10±-0.52 & 0.90±0.02 & 0.76±0.02 \\ 
        RF & TOPOL & G\_offset & 0.93±0.02 & 1.11±0.20 & 2.09±-0.46 & 0.87±0.05 & 0.72±0.05 \\ 
        RF & APFP & G\_offset & 0.93±0.02 & 1.13±0.17 & 2.17±-0.52 & 0.90±0.02 & 0.76±0.04 \\ 
        RF & X-NOISE & G\_offset & 0.92±0.01 & 1.16±0.13 & 2.48±-0.78 & 0.86±0.02 & 0.71±0.02 \\ 
        RF & MolPropsAPFP & G\_FEP & 0.86±0.1 & 1.36±0.4 & 4.0±-1.46 & 0.87±0.05 & 0.73±0.07 \\ 
        RF & MolPropsECFP6 & G\_FEP & 0.86±0.1 & 1.37±0.4 & 4.13±-1.58 & 0.87±0.03 & 0.72±0.06 \\ 
        RF & MolProps & G\_FEP & 0.86±0.1 & 1.37±0.42 & 4.1±-1.53 & 0.86±0.04 & 0.71±0.07 \\ 
        RF & MolPropsTOPOL & G\_FEP & 0.84±0.11 & 1.41±0.37 & 4.56±-1.93 & 0.86±0.04 & 0.71±0.07 \\ 
        RF & TOPOL & G\_FEP & 0.77±0.19 & 1.62±0.43 & 7.11±-3.92 & 0.75±0.16 & 0.60±0.12 \\ 
        RF & APFP & G\_FEP & 0.83±0.13 & 1.87±0.44 & 6.13±-3.12 & 0.64±0.15 & 0.47±0.12 \\ 
        RF & ECFP6 & G\_FEP & 0.37±0.13 & 2.51±0.51 & 14.76±-10.48 & 0.46±0.1 & 0.34±0.09 \\ 
        RF & X-NOISE & G\_FEP & -0.11±0.37 & 2.88±0.33 & 20.32±-15.61 & -0.05±0.47 & -0.04±0.3 \\ 
        MLR & MolPropsTOPOL & G\_offset & 0.96±0.03 & 0.69±0.45 & 1.11±0.31 & 0.94±0.05 & 0.81±0.09 \\ 
        MLR & MolPropsECFP6 & G\_offset & 0.95±0.02 & 0.88±0.29 & 1.45±-0.0 & 0.92±0.05 & 0.77±0.06 \\ 
        MLR & MolPropsAPFP & G\_offset & 0.88±0.05 & 1.7±0.63 & 5.51±-2.45 & 0.85±0.1 & 0.68±0.11 \\ 
        MLR & MolProps & G\_offset & 0.82±0.08 & 1.72±0.75 & 5.99±-2.73 & 0.7±0.13 & 0.54±0.1 \\ 
        MLR & X-NOISE & G\_offset & 0.87±0.03 & 1.74±0.34 & 4.49±-2.08 & 0.77±0.03 & 0.6±0.04 \\ 
        MLR & APFP & G\_offset & 0.1±-0.04 & \textbackslash & \textbackslash & 0.58±0.03 & 0.46±0.05 \\ 
        MLR & TOPOL & G\_offset & 0.21±-0.0 & \textbackslash & \textbackslash & 0.07±-0.02 & 0.03±-0.01 \\ 
        MLR & ECFP6 & G\_offset & 0.04±0.07 & \textbackslash & \textbackslash & 0.07±0.11 & 0.05±0.06 \\ 
        MLR & MolPropsECFP6 & G\_FEP & 0.93±0.05 & 1.12±0.49 & 2.06±-0.17 & 0.85±0.1 & 0.69±0.12 \\ 
        MLR & MolPropsTOPOL & G\_FEP & 0.92±0.06 & 1.17±0.7 & 3.02±-0.59 & 0.89±0.08 & 0.74±0.1 \\ 
        MLR & MolProps & G\_FEP & 0.9±0.17 & 1.37±1.36 & 3.53±-0.02 & 0.86±0.14 & 0.69±0.14 \\ 
        MLR & MolPropsAPFP & G\_FEP & 0.9±0.11 & 1.52±1.27 & 4.88±-1.14 & 0.86±0.11 & 0.7±0.11 \\ 
        MLR & X-NOISE & G\_FEP & -0.27±0.12 & 3.76±1.78 & 31.21±-24.26 & -0.2±0.16 & -0.15±0.09 \\ 
        MLR & TOPOL & G\_FEP & -0.2±-0.01 & \textbackslash & \textbackslash & -0.05±0.0 & -0.03±0.0 \\ 
        MLR & APFP & G\_FEP & -0.2±-0.04 & \textbackslash & \textbackslash & 0.29±0.09 & 0.2±0.06 \\ 
        MLR & ECFP6 & G\_FEP & -0.19±-0.01 & \textbackslash & \textbackslash & -0.13±0.06 & -0.07±0.05 \\ 
        DNN & MolPropsTOPOL & G\_offset & 0.96±0.01 & 0.67±0.17 & 1.09±0.09 & 0.95±0.02 & 0.83±0.04 \\ 
        DNN & TOPOL & G\_offset & 0.96±0.02 & 0.69±0.25 & 1.15±0.13 & 0.93±0.03 & 0.81±0.05 \\ 
        DNN & MolPropsECFP6 & G\_offset & 0.96±0.01 & 0.74±0.19 & 1.18±0.06 & 0.94±0.03 & 0.82±0.05 \\ 
        DNN & MolPropsAPFP & G\_offset & 0.95±0.02 & 0.81±0.24 & 1.45±-0.03 & 0.93±0.02 & 0.79±0.04 \\ 
        DNN & ECFP6 & G\_offset & 0.95±0.01 & 0.88±0.2 & 1.46±-0.09 & 0.93±0.02 & 0.8±0.03 \\ 
        DNN & MolProps & G\_offset & 0.93±0.03 & 1.04±0.34 & 2.22±-0.42 & 0.9±0.05 & 0.74±0.07 \\ 
        DNN & APFP & G\_offset & 0.92±0.02 & 1.11±0.25 & 2.63±-0.77 & 0.9±0.04 & 0.74±0.04 \\ 
        DNN & X-NOISE & G\_offset & 0.92±0.0 & 1.17±0.05 & 2.57±-0.\\
        DNN & MolPropsAPFP & G\_FEP & 0.93±0.1 & 0.95±0.72 & 2.0±0.24 & 0.9±0.11 & 0.76±0.14 \\ 
        DNN & MolProps & G\_FEP & 0.93±0.08 & 1.06±0.69 & 2.18±0.02 & 0.87±0.15 & 0.72±0.18 \\ 
        DNN & MolPropsTOPOL & G\_FEP & 0.92±0.08 & 1.08±0.57 & 2.84±-0.63 & 0.86±0.12 & 0.69±0.13 \\ 
        DNN & MolPropsECFP6 & G\_FEP & 0.92±0.08 & 1.1±0.5 & 2.44±-0.35 & 0.86±0.11 & 0.69±0.12 \\ 
        DNN & TOPOL & G\_FEP & 0.86±0.12 & 1.5±0.61 & 4.28±-1.63 & 0.78±0.15 & 0.6±0.13 \\ 
        DNN & ECFP6 & G\_FEP & 0.69±0.13 & 1.91±0.81 & 8.38±-4.79 & 0.67±0.09 & 0.48±0.08 \\ 
        DNN & APFP & G\_FEP & 0.84±0.12 & 2.19±1.38 & 8.54±-4.16 & 0.62±0.16 & 0.47±0.14 \\ 
        DNN & X-NOISE & G\_FEP & -0.01±0.48 & 2.86±0.05 & 19.86±-15.37 & 0.03±0.48 & 0.03±0.32 \\ 
        \bottomrule        
    \end{tabular}
    \end{adjustbox}
\end{table*}

\clearpage

\section{Hyperparameter space}
We tabulated the parameter optimisation ranges for all independent models (Table \ref{Hyperparameter space}) for ease of viewing. The table shows the optimised parameters corresponding to each model and the total number of parameter searches.
\begin{table*}[htbp]
    \renewcommand{\thetable}{A3}
    \centering
    \caption{Hyperparameter space definition for each model. The hyperparameter with the most accurate model prediction was selected among all parameter combinations, and the model was retrained and tested on the test set under that parameter condition.}
    \label{Hyperparameter space}
    \begin{adjustbox}{width=1\textwidth}
    \begin{ruledtabular}
    \begin{tabular}{cccc}
        \text{ML model} & \text{Hyperparameter} & \text{Range} & \text{Total configurations} \\
        \hline
        \text{SVM} & C & 1e{-}3, 1e{-}2, \ldots, 1e{+}2 & 216 \\
         & $\epsilon$ & 1e{-}3, 1e{-}2, \ldots, 1e{+}2 & \\
         & $\gamma$ & 1e{-}3, 1e{-}2, \ldots, 1e{+}2 & \\
        \text{RF} & \text{NumEstimators} & 1, 2, \ldots, 1000 & 9e{+}4\\
         & \text{MaxDepth} & 1, 2, \ldots, 5 & \\
         & \text{MinSamplesSplit} & 2, 3, \ldots, 10 & \\
         & \text{Bootstrap} & \text{True, False} & \\
        \text{DNN} & \text{ActivationFn} & \text{logistic, tanh, relu} & 3.1e{+}6  \\
         & \text{Solver} & \text{lbfgs, sgd, adam} & \\
         & \text{Layers*} & \makecell[c]{(100, 50), (50, 20), (100, 100, 50),\\(100, 50, 20), (50, 20, 5)}  & \\
         & \text{Adam-}$\beta$ 1 & 0.1, 0.2, \ldots, 0.99 & \\
         & \text{Adam-}$\beta$ 2 & 0.1, 0.2, \ldots, 0.9 & \\
         & \text{Adam-}$\epsilon$ & 10e{-}8, 10e{-}7, \ldots, 10e{-}1 & \\
        \text{MLR} & \text{No hyperparameters to tune.} & & 1 \\
        \text{XGBoost} & \text{NumEstimators} & 1, 2, \ldots, 1000 & \\
        & \text{MaxDepth} & 1, 2, \ldots, 15 & \\
        & \text{MinchildWeight} & 1, 2, \ldots, 10 & \\
        & \text{Subsample} & 0.5, 0.6, \ldots, 1.0 &9e{+}5\\
    \end{tabular}
    \end{ruledtabular}
    \end{adjustbox}
\end{table*}

\clearpage

\section{Feature Analysis}
\subsection{Feature Definations given by XGB Model of Strategy 1}
Specific Definations of Features based on the MolPropsTOPOL dataset given by the XGB(Strategy 1) model (top 11):

GeomShapeIndex: This feature typically represents the geometric shape index of a molecule, reflecting its shape characteristics. The shape of a molecule affects its interactions with solvent molecules, which in turn influences hydration free energy. Molecules with more complex shapes may have larger contact areas and different solvation patterns, making the shape index an important feature.

MOMI-Z: The moment of inertia of the molecule along the Z-axis, as explained earlier, reflects the distribution of molecular mass along the Z-axis. Features related to the molecular geometry and polarity distribution (such as the moment of inertia) may influence the interaction modes of molecules in solvents and thereby affect hydration free energy.

PBF (Partitioned Bond Field): This feature is related to the bond field within a molecule, describing the characteristics of bonds in the molecular structure. The polarity and bonding properties of bonds within a molecule directly affect its interactions in solution, thus influencing hydration free energy.

RNCS (Relative Negative Charged Surface Area): This represents the relative negative charged surface area. The charge distribution on the molecular surface is crucial for its interaction with polar solvents such as water. A higher negative charge surface area is usually associated with stronger solvation effects, which can influence hydration free energy.

TASA (Total Area of Solvent-Accessible Surface Area): Represents the total solvent-accessible surface area. The hydration free energy of a molecule significantly depends on its surface area, as it determines the potential for interaction with the solvent.

RPCG (Relative Positive Charged Surface Area Gradient): This is a feature that describes the gradient of the positively charged surface of a molecule, indicating the variation in the distribution of positive charges on the molecular surface. Similar to negative charges, surface charge distribution has a significant impact on solvation energy.

ATSC5c (Autocorrelation of Topological Structure - Charge at Lag 5): This is a topological autocorrelation descriptor representing the autocorrelation of charges at a lag of 5. It reflects the overall characteristics of molecular charge distribution, which is important in electrolyte solvation processes.

FNSA4 (Fractional Negative Solvent Accessible Surface Area at a Specific Radius): This feature describes the fraction of negative solvent-accessible surface area at a specific radius of the molecule, influencing the solvation behavior in polar solvents.

GATS2s (Geary Autocorrelation of Lag 2, Weighted by I-State): This is a statistical feature that describes the autocorrelation of structural properties (I-state) of a molecule at a specific distance (lag 2), reflecting the atomic interactions within the molecule.

GATS5pe (Geary Autocorrelation of Lag 5, Weighted by Polarizability Eigenvalue): Represents the autocorrelation of polarizability characteristics at lag 5, reflecting the influence of molecular charge and polarity distribution on solvation energy.

FNSA2 (Fractional Negative Solvent Accessible Surface Area at Radius 2): This feature is similar to FNSA4 but represents the fraction of negative solvent-accessible surface area at a different radius, directly affecting the magnitude of hydration free energy.

\subsection{Feature Definations given by XGB Model of Strategy 2}
Specific Definations of Features based on the MolPropsTOPOL dataset given by the XGB(Strategy 2) model (top 10):

AATSC2dv (Average Autocorrelation of Lag 2, Distance Weighted by van der Waals Volume): This is a topological autocorrelation descriptor based on molecular structure, weighted by distance and van der Waals volume (vdW volume) at lag 2. It represents the structural autocorrelation considering van der Waals volume for atomic pairs at a distance of 2 within the molecule. van der Waals volume reflects the size and density of atoms in the molecule. Larger volumes may lead to greater repulsion effects between the molecule and solvent molecules, affecting hydration free energy.

GATS3s (Geary Autocorrelation of Lag 3, Weighted by I-State): This is a Geary autocorrelation descriptor representing the correlation of atomic states (I-state) in a molecular structure at lag 3. The I-state is a feature describing the electronic environment of atoms. This feature can reflect differences and similarities in the electronic environment between atoms within a molecule, influencing the molecule's polarity and its ability to interact with solvent molecules.

PEOE$\_$VSA8 (PEOE Charge Distribution Weighted by van der Waals Surface Area, Bin 8): Uses Gasteiger PEOE charge distribution (Partial Equalization of Orbital Electronegativities) weighted by the van der Waals surface area in the 8th bin. This descriptor partitions the surface area by the magnitude of charges and weights it accordingly. The combination of surface area and charge reflects the polarity and surface properties of the molecule. It directly affects how solvent molecules interact with the molecular surface, thereby influencing hydration free energy.

PBF (Partitioned Bond Field): Describes the bond field characteristics within a molecule, considering the polarity, bond energy, and distribution of bonds within the molecule. It is a comprehensive descriptor. As bond field characteristics are directly related to molecular polarity and electron distribution, it affects the molecule's behavior in the solvent and solvation effects.

ATSC8c (Autocorrelation of Topological Structure - Charge at Lag 8): This is a topological autocorrelation descriptor based on the molecular structure, using charge at a lag of 8. It represents the autocorrelation of charges between atomic pairs that are 8 bonds apart within the molecule. This feature reveals the overall characteristics of the internal charge distribution of a molecule, which may affect interactions between the solvent and the molecule at longer distances, thus impacting hydration free energy.

ATSC5se (Autocorrelation of Topological Structure - Sanderson Electronegativity at Lag 5): This is a topological autocorrelation descriptor based on Sanderson electronegativity, with a lag of 5. It reflects the autocorrelation of electronegativity between atoms that are 5 bonds apart within the molecule. Electronegativity determines the polarity of a molecule and its interactions with solvents. Higher autocorrelation may indicate a greater variation in electronegativity, which could affect the solvation pattern of the molecule.

AATSC5are (Average Autocorrelation of Lag 5, Aromaticity Weighted by Radius Eigenvalue): This is an autocorrelation descriptor based on aromaticity and radius eigenvalue weighting, with a lag of 5. It captures the correlation between aromatic atoms at a distance of 5. The correlation between aromatic atoms can influence the polarity and solvation properties of molecules, especially in polar solvents.

AATSC0i (Average Autocorrelation of Lag 0, Identity Weighted): This is an autocorrelation descriptor with a lag of 0 (i.e., adjacent atomic pairs), representing the autocorrelation weighted by the identity matrix. It is usually used to measure the intrinsic properties of the atoms constituting the molecule rather than their relationships. As a fundamental descriptor, it reflects the intrinsic characteristics of the atoms constituting the molecule, which may indirectly affect the behavior of the molecule in solvents.

GATS3pe (Geary Autocorrelation of Lag 3, Weighted by Polarizability Eigenvalue): Represents the Geary autocorrelation descriptor weighted by polarizability eigenvalue at a lag of 3. It reflects the variation in polarizability within the local region of a molecule. Polarizability affects a molecule's response and interaction with solvents. Higher autocorrelation may indicate uneven distribution of polarizability, affecting solvation effects.

pfp190 (Fingerprint Descriptor, Length 190): This is a path-based molecular fingerprint that represents the presence of atomic paths in the molecule. The pfp190 feature indicates the boolean presence (existence or non-existence) of all atomic paths of length 190 in the molecule. Path length is related to the branching and complexity of the molecule. Longer paths may involve more atoms and bonds, which can affect the interactions between the molecule and solvent and its hydration free energy.

\clearpage
\nocite{*}
\bibliography{aipsamp}

\end{document}